Students' Models of Newton's Second Law in Mechanics and Electromagnetism

Salomon F. Itza-Ortiz, N. Sanjay Rebello and Dean Zollman

*Kansas State University, Department of Physics, Physics Education Research Group,*

*Manhattan KS, 66506*

Abstract

We investigated students' use of Newton's second law in mechanics and electromagnetism contexts by interviewing students in a two-semester calculus-based physics course. We observed that students' responses are consistent with three mental models. These models appeared in mechanics contexts and were transferred to electromagnetism contexts. We developed an inventory to help instructors identify these models and direct students towards the correct one.

I  Introduction

Perhaps the most important topic taught in classical mechanics is Newton's laws. Newton's second law, "F= ma," has preoccupied authors for years.[1-9]  Previous research has suggested that students bring their own understanding of the physical world into the classroom [10] making the teaching and learning of Newton's laws a challenge.[11]  Research also shows that acquisition of knowledge does not happen immediately.  One way the acquisition of knowledge can be conceptualized is that students acquire different understandings of relevant concepts, which coexist and compete with previous informal understandings.[12]  It has been suggested that teachers should be prepared to draw on their students' prior understandings and help them to shape it into one that reflects accepted scientific knowledge.[13]  Cognitive science finds that people tend to organize their experiences and observations into patterns or *mental model(s.* [14]  Thus, by investigating what mental models students use for physics concepts and making this information accessible to teachers we can help them build upon their students' understanding.

In this paper we present part of our research on students' mental models of Newton's second law.  We have probed whether students apply these models consistently across various contexts, through two consecutive semesters of physics, addressing concepts in mechanics and electricity and magnetism.  This kind of research has not been done previously in electricity and magnetism contexts.  We found students used three models for Newton's second law: the Newtonian model ("f=ma"), the Aristotelian model ("f=mv") and a hybrid model that combines the Newtonian and Aristotelian models. In Section II we present definitions for the main terms we use in this work. Section III describes our research methodology, Section IV presents our results and in Section V we discus pedagogical applications and conclusions.

## II Mental Models

We use the term *mental model* to refer to an internal representation, which acts as a structural analogue of situations or processes. Its role is to account for the individuals' reasoning when they try to understand, explain and predict physical world behavior.[15] We consider mental models consist of more fundamental cognitive and knowledge elements, e.g. p-prims [16, 17] or conceptual resources.[18, 19] These elements are assembled consistently and are often called *features* or *aspects* of the model.[20] A few characteristics of mental models include: (1) they consist of propositions, images, rules of procedure and statements as to when and how they are to be used; (2) they may contain contradictory aspects; (3) they may be incomplete; and (4) aspects of a mental model don't have firm boundaries; therefore similar aspects may get confused.[21] The mental models used by the student must be understood in terms of their own internal consistencies, not as "errors" when compared with an expert's model. Mental models used by the students may depend on the context of the problem, e.g. mechanics or electromagnetism. It is possible for a learner to use several different, yet stable and coherent, explanatory elements when tasks are related to different contextual settings pertaining to the same concept.[22]

Several studies of students' intuitive ideas about motion and forces have been conducted. [23-31] Researchers have found that students generally use two ways of thinking about Newton's second law: "Newtonian" and "Aristotelian." Newtonian, when students recognize that *a constant non-zero net force on a body, causes it to increase speed and/or change its direction of motion, (f=ma)* [23, 24] and Aristotelian, when students conclude that *every motion has a cause; an increase in speed is achieved by an increase in force (f=mv)*.[23, 24] These common ways of thinking meet the definition of mental models; therefore, in this paper we will refer to them as the "Newtonian" mental model and the "Aristotelian" mental model. In our research we also found some of our students used aspects from these two main models. We called this new way of thinking a "hybrid" mental model.[20, 32] Students who consistently used any of these three mental models throughout an interview are considered to be in a "pure mental model state." Conversely, students who used more than one mental model during a single interview are considered to be in a "mixed mental model state."[33] In this paper we will also say that the students use a pure mental model or a mixed mental model.

## III Goals and Method

The overarching goal of our research is to develop a multiple-choice instrument, a mental model inventory that allows educators and researchers to probe the mental model states of large numbers of students. To develop such an instrument, we began by exploring the knowledge structures that students use in several contexts through in-depth interviews. Our interviews addressed three main research questions: (1) are the students consistent in their application of models? (2) are students' mental models context dependent? and (3) do particular variables trigger a student's choice of model? We interviewed a cohort of 16 students in a calculus-based physics class six times over the two-semester course sequence. The class operates in a studio format with two one-hour lectures and two two-hour labs integrated with the recitation. The students were volunteers who received monetary compensation. The entire class section consisted of 240 students; majors in different engineering areas, physics and math. About 90% of the students had physics in high school and approximately 5% have taken a physics class at the college level to prepare for the calculus-based physics class, which normally is taken in the sophomore or junior year. About one third of the students were women. The interviews were tape recorded and lasted between 30 and 45 minutes. We proceeded in two phases: Phase I occurred during the first semester when classical mechanics was covered and Phase II occurred during the second semester when electricity and magnetism were covered. In Phase I, eleven men and five women participated in the study. The first interview was conducted prior to instruction on Newton's laws, early in the semester. The second interview was conducted after instruction on Newton's laws and, the third interview was conducted near the end of the semester. The contexts used in the interviews will be discussed in the next section. In Phase II we had ten male and six female participants. Interview 4 (first interview in Phase II) was conducted after instruction on electric fields. Interview 5 was conducted after instruction on magnetic fields, and interview 6 after instruction on induction. Ten of the students --six men and four women-- participated in both phases, completing all six interviews. For the purpose of this paper we limit our analysis to the results from these 10 students. This sample size is considered appropriate for this type of research.[35]

IV Results and Discussion

The protocol for Phase I interviews included two contexts: vertical and horizontal. We used the well-known "Force Concept Inventory" (FCI) [36, 37] as the basis for our research questions. FCI questions 25, 26 and 27 were used for the horizontal context: *a woman exerting a constant horizontal force on a large box. As a result, the box moves across a horizontal floor at a constant speed*. Question 17 was used for the vertical context: *an elevator is being lifted up in an elevator shaft at constant speed by a steel cable*. Each student received a sheet of paper with a figure depicting the problem, a statement of the problem and one or two questions. They were given the opportunity to write notes or make drawings in order to explain their answers. Table 1 presents the protocol question for both contexts and figs. 1(a), (b) show the figures representing the problem.

The three variables explored in these contexts are the magnitude of the force, the mass and the speed of the object. Table 2 shows the students' models through the first semester. An example of a student using the Newtonian model in the horizontal context, interview 1, is student S3. Student S3 responded to question 1, *"it is moving at constant speed, is not accelerating …the woman's force has to overcome friction for the box to start moving, but once moving the forces are equal."* The same student responded to question 2 in the vertical context, part (a), "the steel cable has a force on the elevator… and gravity. Because it is at rest they are equal." part (b) "there is gravity and the force by the cable... the upward force is greater because it is moving up. The force from the cable is larger than it was at rest…gravity is bigger than the force of the cable, gravity does not change…the cable is not exerting as much of an upward force." In the last context S3 is using the Aristotelian model; the elevator is moving up or down, therefore the biggest force must be in that direction. Observe that S3 responded to question 1 in the horizontal context by stating the condition of constant speed, while in response to question 2 in the vertical context there is no reference to this condition. This student used the Newtonian model for the horizontal context, but Aristotelian for the vertical context; therefore, we say the student uses a mixed model state in mechanics contexts. An example of a student who used the hybrid model is S7. S7 responded to question 2 in the horizontal context, "it [the box] will go faster… it would increase its speed to about twice because it is twice the force… it [the speed] increases and then it will become steady to twice the speed [sic]." It is clear that S7 used a combination of Newtonian and Aristotelian models in this response; that is the hybrid model. For the vertical context S7 used the Aristotelian model, thus S7 used a mixed model state in mechanics contexts on the first interview.

The protocol for interview 2 included the same two contexts, horizontal and vertical, plus variants: (1) a woman pushing a box across a horizontal floor (original); (2) a woman pulling a box; (3) a bulldozer pushing a box; (4) a motor pulling a box –with a string; (5) a motor pulling up a box with a cable (originally an elevator); (6) a forklift raising a box vertically, (7) a woman lifting up a box; and (8) a woman standing in a balcony pulling a box with a cable. The variables explored in these scenarios are the same as in interview 1. The interview occurred a few weeks into the term. By this time the students had received two lectures and performed two laboratory activities on Newton's laws. Two students made unexpected changes in their mental models (Table 2). One student (S10) remained in a pure Aristotelian state, while the other (S4) moved from a mixed model state to a pure hybrid state. The other eight students moved to or remained in a pure Newtonian state. These eight students used the line of reasoning "constant speed means no acceleration, that is no net force; then the forces are equal." Students S4 and S10 did not use this logic. Therefore, we believe by failing to recognize this idea, students S4 and S10 did not improve like the other students. S4 did recognize the idea of constant speed as implying equal forces, but when the question about "double the force" or "double the constant speed" was asked, S4 used the Aristotelian model. We noticed that the change in the magnitude of the force and speed as variables induced a student to use the Aristotelian model. This finding is in accordance with other researchers' findings in other contexts.[24] Students used formulas and free body diagrams before attempting to respond to the questions. One of them pointed out that "the formulas give me a guide for answering questions." However, another student showed that having a list of formulas might not be of much help at all. Student S10 used the formula "$f = \frac{1}{2} m v^2$" to respond to our questions. At the time S10 was interviewed the students had covered some energy topics in lecture. S10 drew responses from a wrong formula. We asked the students if they considered the horizontal context problems different from the vertical contexts. The general response was "you just need to think logically, they are the same, forces are forces." We also asked if any particular context was more challenging and what kind of problem was most difficult. The general consensus was *"forces in a horizontal or vertical plane are the same, the only thing [sic]… friction is different from gravity."* And "…inclined planes …you have to work out the components."

For the final interview in Phase I we again used the slightly modified horizontal context --a block on a table pulled using a rope. We also used a modified Atwood's machines --pulleys (Fig. 2). Research has shown that this last context poses difficulties to students.[28] The variables explored in the interviews were again force, mass and speed. Table 2 shows students who used the Newtonian model on interview 2 did not necessarily continue using this model. For instance, student S7 switched from using the pure Newtonian model in interview 2, to using a mixed model (hybrid and Aristotelian). Once again the questions that caused the students to use the Aristotelian model were related to "double the force" and "double the speed." When we asked what is the force used to pull the block at a constant speed, the response was invariably "the force equal [sic] to friction… in opposite direction." The students who continued using the Newtonian model also used the line of reasoning "no acceleration, no net force." Thus, regarding mechanics contexts, the model used by the students depended on the context and not much on when (time frame) the contexts were explored. The idea of "no acceleration, no net force," brought up by students after instruction played an important role in their reasoning.

During the spring term we conducted Phase II of our research. The second semester of the course traditionally starts with electrostatic concepts and then moves to magnetism.[34] The fourth interview (first in Phase II) in our research was conducted after instruction on electric fields. Our interest was not the students' knowledge on electric fields, but their use of Newton's second law.  As the basis for the electric field context we used question 10 from the Conceptual Survey in Electricity and Magnetism (CSEM) [38] with slight modifications:  *A positively charged sphere is released from rest in a region with a uniform electric field* (Fig. 3).  The questions focused on five variables -- the magnitude of the field, the mass, the magnitude of the charge, its sign and the speed.  A second problem statement within this context was, *A positively charged sphere moves at a constant speed in a uniform electric field.*  Table 3 shows the questions asked in this context.  For the second problem we added one more question: If the speed of the sphere is twice the original, how does the E field change?  Following our protocol from Phase I, all students received a sheet of paper with a figure, the statement of the problem and two questions.  To continue exploring whether the students' models change with time, besides context, we gave the students two problems relating to Phase I -- the horizontal context FCI #25 [36, 37] and a box sliding from an incline plane.  The question was, describe the force(s) acting on the box.  We found that two months after the last interview students who used the Newtonian model continued to do so.  They still used the line "no acceleration, no net force."  Student S1 responded to question 1, in the case where the sphere moves at constant speed:  "My instincts from mechanics… say constant speed is net force equal zero… so there is a force from E to the right [looking at the figure on the paper] some force should be pushing to the left … I am not sure what force [referring to the force to the left]."  For this student the E field context was asked before the FCI question, so no triggering by mechanics contexts can be claimed.  Table 4 shows the models used by students during Phase II.  Thus, though electric fields contexts are not frequently related to the use of Newton's laws, students did use the Newtonian model, when appropriate, if they recalled basic principles.  A few of the students used the equation "F =qE" and drew arrows before responding.  All students answered that the force is proportional to the magnitude of the charge and that of the electric field.  When we asked question 4, double the mass of the sphere, we did not give clues as to the size of the sphere.  A few of the students commented that in "this class" (referring to their physics lecture) "mass is negligible."  We do not know if this idea came from their instructor or through solving problems.  However other students did consider the mass of the sphere and gave appropriate answers by using "F=qE = ma."

      For interview 5 we used a similar protocol to the one for electric fields, but instead of electric field we used uniform magnetic fields B.  In the first problem the charged sphere was set at rest.  In the second problem the sphere moves at constant speed with a horizontal direction into an area with a magnetic field with the same direction; and the third problem differs from the second in that the direction of the magnetic field is perpendicular to the velocity of the sphere.  The interviews were conducted after instruction on B fields.  To identify what model students used, we took into account only the way they used forces.  When the sphere was set at rest, three of the students responded by considering an E field instead of a B field; that is, they explained

that there is a force in the direction of the field, instead of no force since the sphere was set at rest. Two of these students changed their answers later when we brought up the second problem statement; the idea that the sphere must have a speed to have force acting upon it came back to them. We noticed that students again used formulae before answering the questions. For example, student S8's response to the question: describe the force(s) acting on the charged sphere (when the sphere is at rest) was "the speed is zero, and… F=qVxB… the magnetic field cross the speed is equal zero…so no force is acting on the charged sphere. no force is no acceleration, the sphere does not move." The line of reasoning "no force-no acceleration-constant speed" is present. Student S9 also responded to this same question using the same line of reasoning. However, when the same question was asked in the case where the magnetic field's direction is perpendicular to the velocity's direction, student S9 responded: "the sphere cross the direction of B [sic]…B in the y direction when sphere enters the field region the force is out of the page [gestures of using right hand rule]. The motion of the sphere is F=qVB…. As the particle enters it will come out… it will come out at constant speed." Student S9 is one of the students who has used the line "no force-no acceleration-constant speed," but failed to use it correctly in this context. Again, the context does influence the model students use (Table 4).

In the final interview we explored the context of induction. We used two typical problems. The first one was *A loop is pulled, with constant speed, out of a region with a uniform magnetic field B*. The second was *A rod of length L moves at constant speed on two rails in a uniform magnetic field B* [34, p 718,740]. We asked two questions for each problem. The first question was to describe the forces acting on the loop (or rod). The second question related forces and motion of the loop (or rod) when pulling was stopped (or the B field is turned off). In addition we asked the students about how their thinking on these problems related to their first semester course, especially about Newton's second law. Table 4 shows that none of the students used the Aristotelian model on the induction context. We could claim that the students moved forward on their understanding since the use of a hybrid model implies the use of aspects of the Newtonian model. However, students S7 and S10 switched from the Newtonian to the hybrid model. To find what might have caused this change we carefully reviewed student S10's transcript. S10 responded to the question related to turning off the B field on the rod problem "well, there is no longer a B field, since the force F = I L x B is zero… the rod should come to a stop." S10's response to the extra questions was: "I do not remember using Newton's second law…uh... F=ma [respect to the motion of the rod]. …If there is a force there must be an acceleration, and if there is one force there is no net force." We concluded that student S10 forgot about the meaning of "net force" or perhaps has not understood its meaning. S10 used the Newtonian model only in the B field context.

# V Teaching Implications and Conclusions

We found that students solved problems related to Newton's second law using two main mental models labeled "Newtonian," and "Aristotelian." They also used a third model which we labeled hybrid model. When a student used only one of these three models in an interview it is said that the student used a pure mental model. When the student used more than one model, it is said that the student used a mixed model. Other authors say the students are in a pure or mixed model state.[32, 33] In tables 2 and 4 we follow the progress of 10 of the students through different contexts and time. In the first semester, before instruction, only one student (S5) used a pure Newtonian model. After instruction six more students used the pure Newtonian model. Perhaps the most important idea that produced this change was the line of reasoning: "constant speed means no acceleration, that is no net force; then forces are equal." The Aristotelian and hybrid models were mostly triggered by questions relating to double the force, double the mass or double the speed; perhaps because of the word "double" which indicated proportionality. Students transferred their models to the second semester. Seven of the students used a pure Newtonian model on the first interview of Phase II. This was expected since the students had received instruction on Newton's second law. It is also important to note that in electromagnetism contexts, when students are faced with abstract contexts, they are more likely to base their responses on instruction and not on intuitive reasoning as they might do in mechanics contexts. The understanding of basic concepts like "vector" and vocabulary such as "net force" also has a role on the model a student might use. It is important that the instructor reviews concepts and vocabulary for Newton's second law. Students might be using incorrect models because of the misunderstanding of the background and not because of the misunderstanding of Newton's second law. For the transfer of models from electromagnetism to mechanics contexts, it is important how electromagnetism contexts are introduced. Some students considered that the transfer of Newton's laws from mechanics to electromagnetism contexts was not clear. They pointed out that in most of their homework problems mass is negligible, things are small and there are many approximations.

With the results of these six interviews we have developed a mental model inventory (multiple-choice instrument). This inventory is a tool for instructors to help them determine the models students use; then the instructor can use the information to tailor a class to correct the students' models. The inventory consists of five surveys that address the same contexts as our interviews. Each survey has five to eight questions with four or five options. The options refer to the possible models students might use. The number of questions is short to facilitate use of the surveys in class. If the instructor uses a personal response system, (s)he could obtain immediate feedback on the model students used. Survey 1 focuses on the contexts of interview 1, horizontal vertical contexts. Survey 2 focuses on Atwood's machines, survey 3 on electric fields, survey 4 on magnetic fields and survey 5 on induction. The surveys can also be used as an assessment tool and possibly to continue research on mental models. The entire inventory is available online as portable document format (PDF) files at http://web.phys.ksu.edu, under "Research in Education," and then "Model-Based Assessment."

## Acknowledgments

This work was supported, in part, by the National Science Foundation under grant # REC-0087788. We thank Zdeslav Hrepic and Erin Itza for comments and suggestions.

TABLES

Table 1

| Horizontal context | Vertical context |
|---|---|
| 1.-How does the force exerted by the woman compare with other forces acting on the box? (FCI #25) | 1.-How does the force exerted by the cable compare with other forces acting on the elevator? (FCI #17) |
| 2.-How will the speed change if her force is doubled? (FCI # 26) | 2.-What is(are) the force(s) acting on the elevator when (a) it is held at rest? (b) it is moving up or down at constant speed |
| 3.-What force is needed to double the speed? | 3.-What is the force if the speed is doubled? |
| 4.-What force is needed to steadily increase the speed? | 4.-What is the force if the speed is steadily increasing? |
| 5.-What happens if she stopped pushing? (FCI # 27) | 5.-How does the speed change if the force is doubled? |
| 6.-What would happen if she exerts the same force on two boxes, one of top of the other? | 6.-What force is needed to move an elevator twice as massive at the same speed? |

Table 2

| Interview / Context | S1 | S2 | S3 | S4 | S5 | S6 | S7 | S8 | S9 | S10 |
|---|---|---|---|---|---|---|---|---|---|---|
| *1/ Horizontal / Vertical* | H A | H N | N A | H N | N N | N H | H A | A A | A A | A A |
| **2/ Horizontal / Vertical** | N N | N N | N N | H H | N N | N N | N N | N N | N N | A A |
| **3/ Horizontal / Pulleys** | N N | N N | N N | H H | N N | N N | H A | N N | N H | H A |

Table 3

| Electric field context | |
|---|---|
| 1.-Describe the force(s) acting on the charged sphere. | 4.-If the mass of the sphere is doubled, describe its motion. |
| 2.-Does the motion of the sphere change if the magnitude of the E field is doubled? | 5.-How would the motion of the sphere change if the direction of the E field were reversed? |
| 3.-If the charge of the sphere is doubled, describe its motion. | 6.-How would the motion of the sphere change if the E field were turned off? |

Table 4

| Interview / Context | S1 | S2 | S3 | S4 | S5 | S6 | S7 | S8 | S9 | S10 |
|---|---|---|---|---|---|---|---|---|---|---|
| *4/ Horizontal / E field* | N<br>N | N<br>N | N<br>N | H<br>H | N<br>N | N<br>N | A<br>H | N<br>N | N<br>N | A<br>A |
| **5/ B field** | N | N | N | A | N | N | N | N | H | N |
| **6/ Induction** | N | N | N | H | N | N | H | N | N | H |

FIGURES

Figure 1 (a), (b)

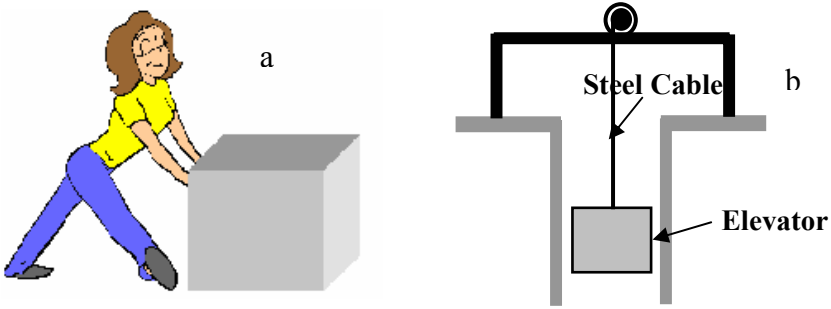

Figure 2

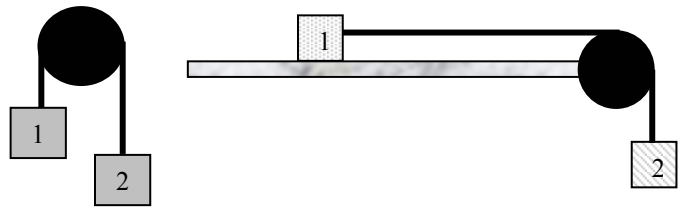

Figure 3

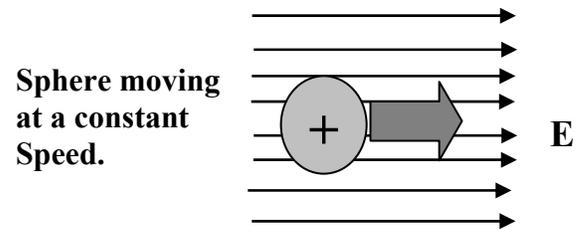

CAPTIONS

Table 1 Questions asked to students in interview 1 of Phase I, mechanics.

Table 2: Students' models in interviews, Phase I. The 'A' indicates "Aristotelian" model; 'N' indicates "Newtonian" model and H indicates Hybrid model.

Table 3. Questions asked to students in interview 1 of Phase II, electromagnetism.

Table 4: Students' models in interviews, Phase II. The 'A' indicates "Aristotelian" model; 'N' indicates "Newtonian" model and H indicates Hybrid model.

Figure 1 (a) From FCI questions 25-27, a woman pushing a box. (b) From FCI question 17, an elevator being lifted up.

Figure 2. Pulley system used in protocol for interview 3 (Atwood's machine). Masses 1 and 2 are identical.

Figure 3. A charged sphere moving at constant speed in a uniform electric field E. Modification from CSEM question 10.